# Simulation of the band structure of InAs/GaSb type II superlattices utilizing multiple energy band theories

**Shuiliu Fang[1], Ruiting Hao[2*], Longgang Zhang[1], Jie Guo[1], Wuming Liu[3,4*]**

[1]Yunnan Key Lab of Opto-electronic Information Technology, School of Physics and Electronic Information, Yunnan Normal University, Kunming, Yunnan Province 650092, China

[2]School of Energy and Environment Science, Yunnan Normal University, Kunming, Yunnan Province 650092, China

[3]Beijing National Laboratory for Condensed Matter Physics, Institute of Physics, Chinese Academy of Sciences, Beijing 100190, China

[4]Songshan Lake Materials Laboratory, Dongguan, Guangdong Province 523808, China

**\* Correspondence:**
Ruiting Hao
ruitinghao@semi.ac.cn

Wuming Liu
wliu@iphy.ac.cn

**Abstract**

Antimonide type II superlattices is expected to overtake HgCdTe as the preferred materials for infrared detection due to their excellent photoelectric properties and flexible and adjustable band structures. Among these compounds, InAs/GaSb type II superlattices represents the most commonly studied materials. However, the sophisticated physics associated with the antimonide-based bandgap engineering concept started at the beginning of 1990s gave a new impact and interest in development of infrared detector structures within academic and national laboratories. InAs/GaSb superlattices is a type II disconnected band structure with electrons and holes confined in the InAs and GaSb layers, respectively. The electron micro-band and hole micro-band can be regulated separately by adjusting the InAs and GaSb layers, which facilitates the design of superlattice structures and maximizes the amount of energy band offset. In recent years, both domestic and foreign researchers have made many attempts to quickly and accurately predict the bandgaps of superlattice materials before superlattice materials grow. These works constituted a theoretical basis for the effective utilization of the InAs/GaSb system in material optimization and designing new SL structures; they also provided an opportunity for the preparation and rapid development of InAs/GaSb T2SLs. In this paper, we systematically review several widely used methods for simulating superlattice band structures, including the k·p perturbation method, envelope function approximation, empirical pseudopotential method, empirical tight-binding method, and first-principles calculations. With the limitations of different theoretical methods proposed, the simulation methods have been modified and developed to obtain reliable InAs/GaSb SL energy band calculation results. The objective of this work is to provide a reference for designing InAs/GaSb type II superlattice band structures.

**Keywords:** InAs/GaSb, Type II Superlattice, Band Structure, Bandgap, Theoretical Calculation and Simulation.

## 1 Introduction

Infrared detectors and lasers are employed in both military and civil applications. As more advantages of type II superlattice (T2SL) materials are discovered, T2SL infrared lasers and detectors containing InAs, GaSb, and AlSb compounds are expected to be more widely used after HgCdTe and multi-quantum wells[1]. Unlike HgCdTe, the antimonide type II SLs exhibit a high level of reproducibility and maneuverability, large area uniformity, and low Auger recombination rates, which means that T2SL infrared lasers and detectors have lower dark currents, high-temperature operational characteristics [2], and distinct advantages in some application scenarios. For a long period, many laboratories worldwide have invested manpower and resources to perform theoretical simulations in the field of energy bands and achieved significant progress.

T2SLs were originally proposed by Esaki and Tsu in 1970 [3]. They represent periodic structures composed of two or more semiconductor layers of III–V materials with a type II band alignment and lattice constant of approximately 6.1 Å. Theoretical simulations of T2SLs related to energy band engineering applications have been initiated in the early 1990s [4]. Among these materials, InAs/GaSb SLs exhibit high flexibility in terms of bandgap adjustment and heterostructural design in the mid-wave infrared (MWIR) and long-wave infrared (LWIR) regions due to their unique properties including the lower position of the InAs conduction band than that of the GaSb valence band [5], low Auger recombination rate [6], and large effective mass [7], which attracted considerable attention from researchers [8]. However, the performance of T2SL devices is significantly lower than theoretical predictions. Moreover, InAs/GaSb SLs have several disadvantages, including small carrier lifetimes, high dark currents, and low quantum efficiencies, which are attributed to the lack of a clear understanding of their band structure and topology [9]. Therefore, simulating SL band structures is the most important step in designing SL infrared lasers and detectors [10]. To achieve this goal, it is necessary to select appropriate theoretical methods, establish accurate device models, and propose new design improvement strategies by analyzing the physical properties of T2SL materials and/or devices. For this purpose, multiple studies on the material growth, electronic properties, and structural design of InAs/GaSb SLs have been conducted.

This article outlines the energy band structure of InAs/GaSb SLs, describes in detail several theoretical simulation methods of solid energy band commonly used for studying SL energy band structures, such as the k · p perturbation method, envelope function approximation, empirical pseudopotential method, empirical tight-binding method, density functional theory, and many-body terturbation theory. In the last section, it discusses the bottleneck problems and development trend of SL energy band simulation techniques.

## 2 Type II InAs/GaSb SL electronic band structure

InAs/GaSb SLs was originally developed by Sai-Halasz et al. in 1977 [11]. It is a periodic structure formed by InAs (a=6.0584Å) and GaSb (a=6.0959Å) grown alternately for several cycles. And it is worth mentioning that the superlattice bandgap is determined by the energy difference between the electron miniband $E_1$ and the first heavy hole miniband $HH_1$ at the center of the Brillouin zone. The bottom of the InAs conduction band is much lower than the top of the GaSb valence band, which corresponds to a staggered type II band structure (Fig. 1(a)) with electrons and holes confined in the InAs and GaSb layers, respectively. There is also a mutual coupling between the wavefunctions in the quantum well and the transition only occurs in the spatial region where the wave functions overlap, which broadens the electron and hole sublevels to form an energy band with a certain width (Fig. 1(b)). It has been confirmed in the research model of Becer, et al[12]. The separation of electrons and holes in the real space not only effectively suppresses the Auger recombination of carriers, but also enables the independent adjustments of the electron and hole potential wells to achieve continuous light absorption in the wavelength range of 2–30 μm [13].

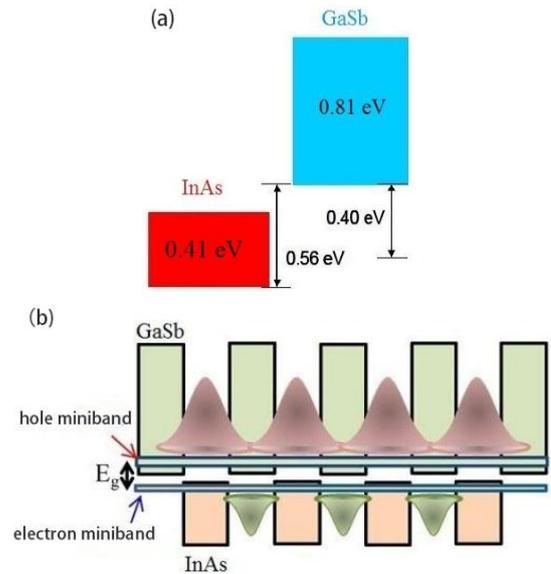

Fig. 1. (a) In the InAs/GaSb superlattice, the conduction band of InAs is about 0.15eV lower than the valence band of GaSb, and its heterojunction forms an off-type energy band. (b) InAs/GaSb superlattice energy band structure is a staggered type II band structure, causing electrons and holes to be confined in the InAs layer and GaSb layer, respectively.

The use of T2SLs for the fabrication of lasers and detectors depends [14] not only on the ability to grow a perfect periodic crystal structure, but also on the material band-gap design [15]. During the selection of a device cut-off wavelength, the SL bandgap can be theoretically adjusted by varying the thicknesses of the InAs and GaSb layers and thus the degree of overlap of SL electronic wavefunctions. Delmas, et al. applied simulation tools to model and design high-performance InAs/GaSb T2SLs infrared detectors, showing that the SL design can improve overall device performances [16]. Today, many laboratories also showed that the experimental absorption spectra of the MWIR and LWIR InAs/GaSb and InAs/InAsSb T2SLs could be accurately simulated (Fig. 2). The emergence of band gap engineering has promoted the development of infrared detector structures.

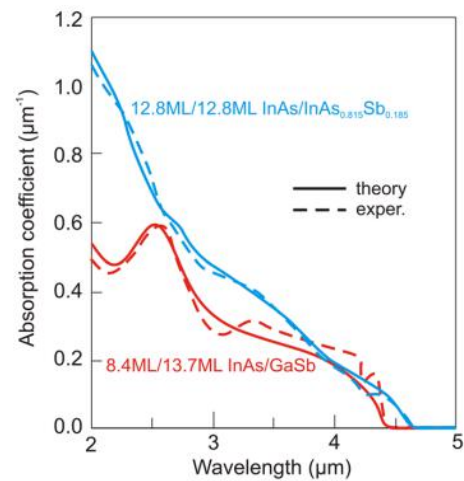

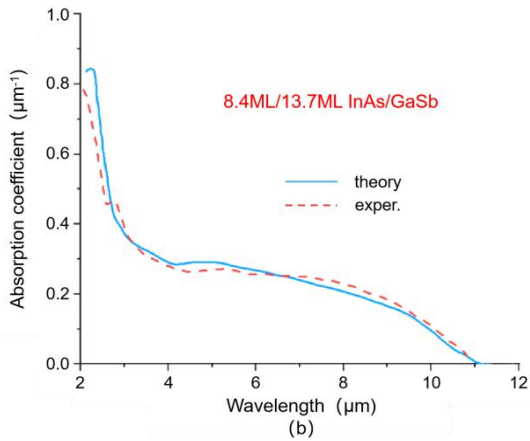

Fig. 2. Measured and calculated absorption spectra of the (a) the mid-wave infrared 8.4 ML InAs/13.7 ML GaSb and 12.8 ML InAs/12.8 ML InAsSb and (b) the long-wave infrared 8.4 ML InAs/13.7 ML GaSb T2SLs. The strong peak below 3 μm was due to the transition from $HH_2$ to $C_1$ at the boundary of the Brillouin zone. The experimental spectrum and theoretical calculation are in good agreement.

In other words, to better understand the properties of T2SLs, their band structure must first theoretically simulated. The theoretical methods currently used for this purpose include the k·p perturbation method, envelope function approximation (EFA), empirical pseudopotential method (EPM), empirical tight-binding method (ETBM), and first-principles calculations.

## 3 Theoretical simulation methods

### 3.1 K·p Perturbation Method

The k·p model is called a standard model because its calculation results are easily explained by the physical theory. In addition, the related calculation procedure is relatively simple and requires inputting a few parameters to solve the Schrödinger equation by changing the electron potential energy, while the SL band structure can be obtained from the wavefunction. The k·p perturbation method is based on the envelope function and effective mass approximations. It was introduced by Bardeen [17] and Seitz [18], developed by Bastard [19] in 1988, and applied to study the T2SL structures of infrared photoelectric detectors by Smith and Mailhiot [20].

After Read and Shockley extended the k·p perturbation method to an eight-fold degeneracy, the eight-band k·p matrix was used for the InAs/GaSb SL energy band simulation [21]. Klipstein derived an 8×8 envelope function Hamiltonian for the SL structure of $\Gamma_{15v}$ and $\Gamma_{1c}$ in 2010 [22]. Subsequently, Livneh et al. reduced the computational error by introducing an eight-band Hamiltonian to eliminate the terms with energies below the expected accuracy level [23]. Using this approach, the same researchers ultimately decreased the number of unknown fitting parameters to six. Afterwards, they applied the eight-band k·p model for fitting the absorption spectra of InAs/GaSb SLs in the wavelength region of 4.3–10.5 μm at temperatures of 77 and 300 K to determine the six Luttinger and interfacial parameters. The fitted Luttinger parameters were very close to those originally proposed by Lawaetz [24], which indicated that the eight-band k·p model retained the high calculation accuracy. Finally, they used this model to predict the wavelengths of more than 30 SLs. The obtained photoluminescence (PL) spectra demonstrated that most errors did not exceed 0.3 μm, the maximum error was 0.6 μm, and the corresponding layer thickness error was less than 0.4 ML.

In 2010, Rejeb et al. simulated the structure of short-period InAs/GaSb/InSb SLs on GaSb substrates using the eight-band k·p method and plotted the fundamental bandgaps of the mutated and separate interfaces as functions of the period number N [25]. The obtained results indicated that in the case of interfacial interactions, the asymmetric interfacial segregation could lead to a bandgap reduction of approximately 30%. In 2012, Qiao et al. developed a band structure model using an eight-band k·p method with Dirichlet and periodic boundary conditions that included the actual interfacial layers, which was achieved by changing the previous way of adjusting valence band offset values or using potential gradient profiles [26]. Klipstein et al. used this k·p model to simulate InAs/GaSb SLs in 2014, which also considered interfacial effects and bandgap bending [27]. At a temperature of 10 K, the calculated bandgap was consistent with the PL peak energy (Fig. 3). If the interfacial matrix was ignored, the calculated 8.4 ML InAs/13.7 ML GaSb MWIR T2SL bandgap would exhibit a blue shift of 0.75 μm, and the 14.4 ML InAs/7.2 ML GaSb LWIR T2SL bandgap would produce a blue shift of 4.5 μm. In 2019, Delmas, et al. applied the eight-band k·p method to calculate the cut-off wavelengths of InAs/GaSb SLs with four different period lengths: 10/4, 12/4, 14/4, and 17/4, and compared the results with or without considering the interface matrix $H_{IF}$ and the InSb interface [28]. Fig. 4 showed that there was good agreement between simulation and experiment, and the error was within deviation range. If both the $H_{IF}$ and the InSb interface are not considered, the model didn't predict the measured cut-off wavelength and underestimated it. Therefore, interfacial effects cannot be ignored when the eight-band k·p model is employed for calculating the SL energy band structure.

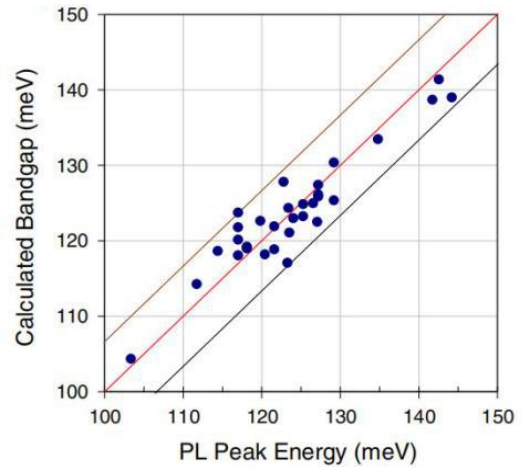

Fig. 3. Calculated bandgaps and PL peak energies measured at 10 K for more than 30 InAs/GaSb T2SLs spanning from the MWIR to LWIR wavelengths. The thin lines denote the deviations from the ideal behavior (thick central line) at 77 K.

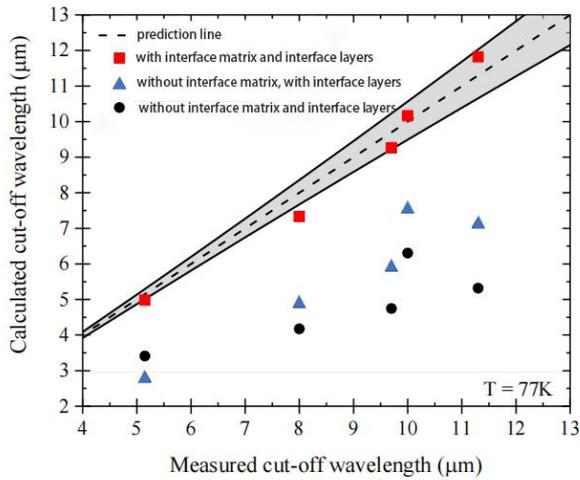

Fig. 4. Comparison between the calculated and measured cut-off wavelengths at 77K for the different SL periods (red squares) along with the ideal prediction line (dashed line). Cut-off wavelengths calculated without the interface matrix (triangles) and considering neither the interface matrix nor the InSb layers (circles) are also plotted for comparison. The deviation in the predicted is represented by the solid lines and the grey area. The 10/4 SL was under compressive strain on GaSb with a large lattice mismatch and error.

The above-mentioned k·p method only considers the eight-fold degeneracy. Hence, without taking into account interfacial effects, the eight-band k·p method overestimates the SL bandgap by neglecting the interactions between the low-energy band of the GaSb layer and the high-energy band of the InAs layer. However, a more accurate bandgap value can be calculated by a modified k·p method described below.

In 2002, Vinter [29] proposed an 18-band k·p method to describe the SL wavefunction. The very small differences between the theoretical and experimental energy values are shown in Fig. 5. In 2015, Imbert et al. used the 18-band k·p model to simulate three devices with cut-off wavelengths of 5 μm but different lattice periods, which corresponded to the thickness ratios R between the InAs and GaSb layers equal to 0.5, 1, and 2 [30]. The relationship between the bandgap energy and the period thickness determined for the symmetric structure (R = 1) (Fig. 6) demonstrated a good agreement between the measured bandgap values and those calculated by the 18-band k·p model [31].

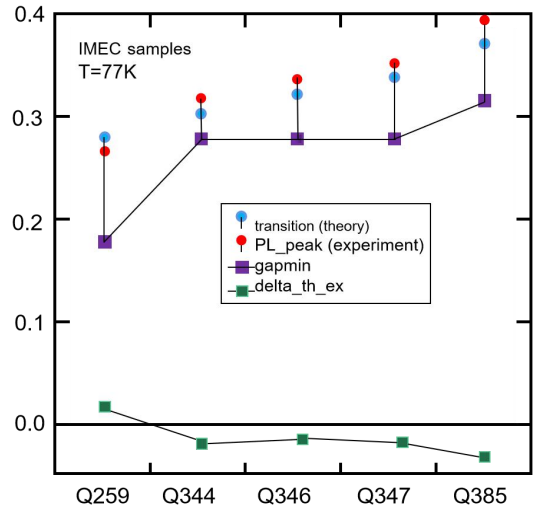

Fig. 5. Theoretical $E_1 - H_1$ transition and experimental PL energies. Here, "gapmin" represents the difference between the bottom of the conduction band and the top of the valence band. The differences between the theoretical and experimental values (marked "delta_th_ex") are very small despite the considerably variations of the confinement energy.

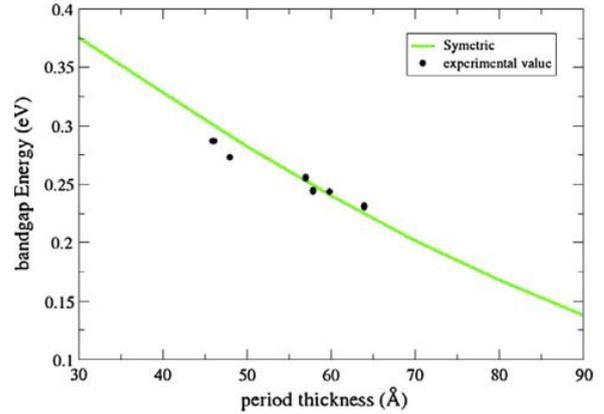

Fig. 6. Calculated (solid line) and measured (black symbols) bandgaps the of symmetrical InAs/GaSb SL (R = 1) on a GaSb substrate at 77 K plotted as functions of the SL period thickness.

In 2018, Machowska−Podsiadlo et al. used the eight-band k·p method to study the effects of temperature, band offset energy, strain, and interface on the edge of the SL absorption band and relative contributions of these parameters [32]. The results of five simulations conducted for the 8 ML InAs/8 ML GaSb SL are presented in Fig. 7. They show that the SL bandgap varies slightly in the temperature range of 0−77 K, which is in good agreement with previously obtained data [33; 34]. Subsequently, the authors studied m/8 and 8/n SLs more comprehensively. They found that regardless of the GaSb layer thickness, the effect of strain on the absorption edge of the 8/n SLs remained the same; however, for the m/8 SLs, this effect was more pronounced at larger thicknesses of the InAs layers. Moreover, the bandgap energy computed at $E_{offset}$ = 140 meV was 7−9 meV higher than that calculated at $E_{offset}$ = 150 meV, which also agreed with the experimental value [35]. The authors concluded that the leap energy level in SLs mainly depended on the interface type. The results of the last study can help design more complex InAs/GaSb SL structures.

In 2019, Jeffrey established an accurate 8-band k·p model by optimizing the parameters in the experimental data, and calculated the band structure of the InAs/GaSb superlattice[36]. As shown in Fig.8, the calculated absorption spectrum shape of 4-10 um was consistent with the experimental data. The calculated absorption cut-off values were consistent, confirming that the position of the quasi-Fermi level can be set appropriately. Kim reported the modeling results of optical and electrical characteristics of mid-wave infrared InAs/GaSb T2SLS [37]. They used the improved k·p model to calculate the bandgaps of SLs as a function of the thickness of InAs and GaSb (Fig. 9), which proved the fact that the bandgap can be effectively adjusted by adjusting the thickness of InAs or GaSb. Du, et al.[38] investigated the electronic band structure of InAs/GaSb SLs by 8-band k·p method and compared it with density functional theory (DFT) [39], empirical pseudopotential method (EPM) [40] and experimental results [41]. As shown in Fig. 10, it verified the accuracy of the 8-band k·p theoretical model. In the same year, Cui, et al. designed an M-structure T2SL detector with a cut-off wavelength of 10.5 μm based on the eight-band k·p model and studied its photoelectric performance [42]. In 2021, Mukherjee, et al.[43] simulated the 8ML InAs/8ML GaSb T2SL band structure. The calculated bandgap was 0.27 eV within the k·p model under the envelope function approximation at 77 K, and the corresponding cut-off wavelength was 4.59 μm. It was in good agreement with the experimental bandgap in the range of 0.269-0.275 eV.

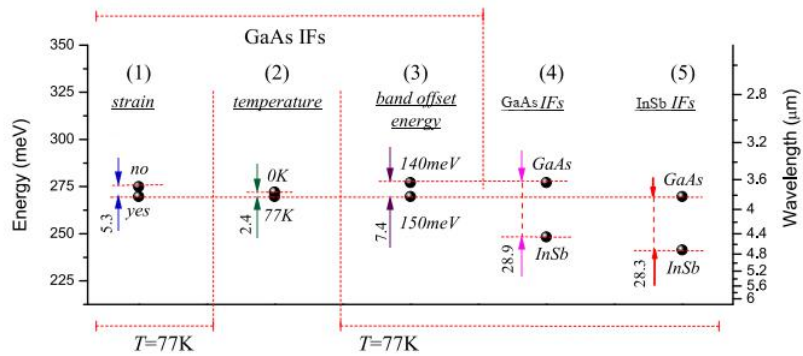

Fig. 7. Effective bandgaps and the corresponding cut-off wavelengths calculated for the 8/8 ML SL structure. Simulation data were obtained for the following cases: (1) when strain effects were taken into account/neglected ("yes/no", respectively), (2) for two temperature values ("0K" and "77K"), (3) for two Eoffset energy values ("140meV" and "150meV"), (4) and (5) for two types of interfaces in the SL ("GaAs" and "InSb"), and for Eoffset energy values equal to (4) 140 meV and (5) 150 meV.

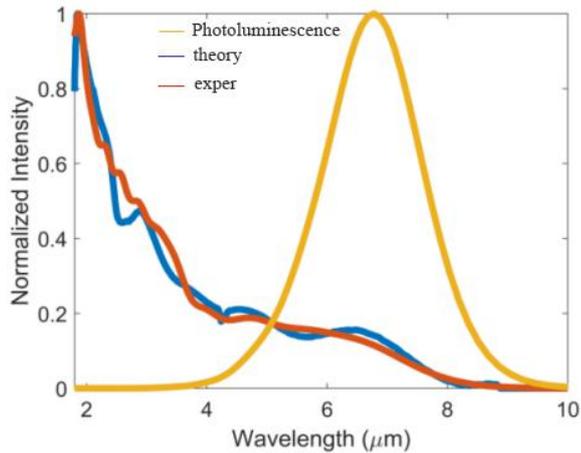

Fig. 8. The results of optimizing parameters to a 13/7 InAs/GaSb superlattice. The experimental absorption spectrum (blue), calculated absorption spectrum (red), and Photoluminescence spectrum (yellow) are shown.

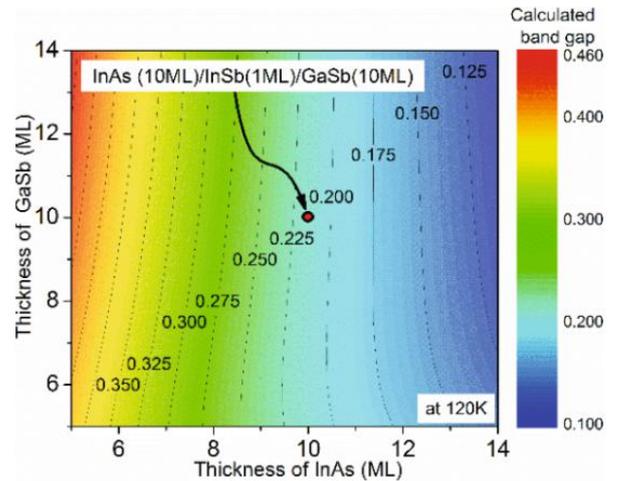

Fig. 9. The calculated bandgaps using the modified K·P model. The x and y axes represent the number of InAs and GaSb monolayer, respectively. The calculated bandgap of 10ML InAs/1ML InSb/10ML GaSb T2SL is 0.2 eV at 120 K. And as the thickness of InAs increases while keeping the thickness of the GaSb layer constant, the cut-off wavelength of the InAs/GaSb T2SL also increases.

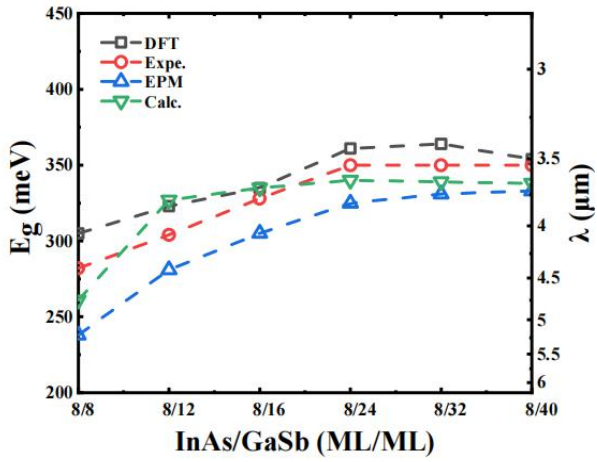

Fig. 10. Comparison of bandgaps calculated by k · p theory with DFT, EPM and experimental results. The black square are the bandgap $E_g$ calculated by density functional theory (DFT), the red circle is the bandgap of experimental results, the bule upward triangle is the bandgap calculated by empirical pseudopotential method (EPM), and the green downward triangle is calculated results by k · p method in this work.

In summary, the k·p standard model is intuitive and easy to explain. It is an effective and widely used method for determining the energy band structure near the bottom of the conduction band and the top of the valence band in semiconductors. However, this technique utilizes the effective mass approximation while ignoring atomic parameters and cannot accurately describe the electronic structure of a short-period SL [44], and it may provide similar result in MWIR and LWIR. Therefore, it is mainly used to simulate quantum wells, SL quantum dots, and long-period materials, while a new k·p theoretical method suitable for short-period T2SLs must be developed separately. In addition, the k·p model contradicts the low-momentum hypothesis of the k·p theory in many applications, which is controversial on a theoretical basis.

### 3.2 Envelope Function Approximation

In the early studies on InAs/GaSb SLs, the numerical accuracy of the standard EFA algorithm was considerably increased [45; 46]. The early model still did not consider the influence of interface on the SL band structure. As a result, the bandgaps of the first conduction band and heavy hole microstrip were significantly overestimated [47].

In 2004, Szmulowicz et al. proposed an improved 8×8 EFA method, which considered the effects of anisotropy and interfacial couplings for non-coatomic SLs [35]. In their study, a 44.16 Å GaSb/55.46 Å InAs SL structure was combined with an InSb interface. The simulated bandgap was 111.4 meV, which was closer to the experimental value than the magnitude of 116.9 meV computed by the 14 × 14 k · p perturbation method. Therefore, the modified EFA method including interfacial effects can provide more accurate simulation results for SL systems. In 2005, J. et al. studied m ML InAs/m ML GaSb SLs with and without additional interfacial potentials using the same model [47]. As shown in Fig. 11, the fitted data of the modified EFA model considering strong perturbations at the interface are in good agreement with the experimental values. Subsequently, Szmulowicz et al. selected a smoother InSb interface with higher carrier mobility. Using the modified 8 × 8 EFA model [48], they designed an MWIR superlattice with a bandgap of 310 meV and cut-off wavelength of 4 μm. The calculated bandgaps of the 23.9 Å GaSb/20.4 Å InAs SL with a shorter period and 9.9 Å GaSb/11.4 Å InAs SL with a longer period were 304.2 and 313.8 meV, respectively. The bandgap values computed without taking into account the interfacial effect were equal to 391.5 and 514.6 meV, respectively.

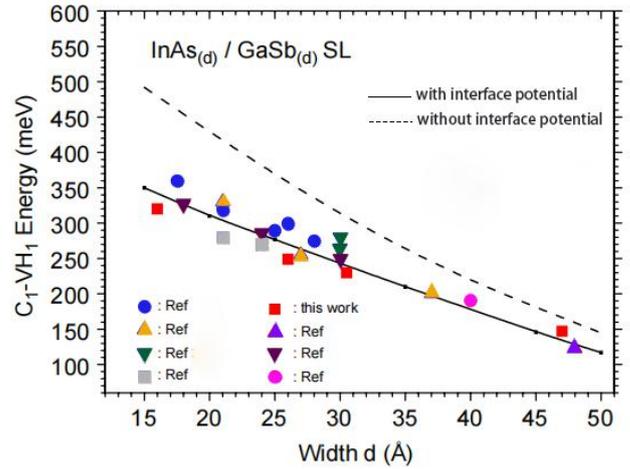

Fig. 11. Evolution of the $C_1-VH_1$ fundamental interminiband transition energy of the InAs(m)/GaSb(m) SL structure with the layer thickness: theoretical calculations conducted with (solid line) and without (dashed line) an additional interface potential; experimental data extracted from PL measurements in this work (solid square) and from other works.

In the same year, Haugan et al. designed 4-μm short-period InAs/GaSb SLs using the EFA model [49]. In Fig. 12, the measured peak position consistently maintains a constant level of 3.757 ± 0.10 μm (330 ± 10 meV) despite the large thickness variations from 50.2 to 21.2 Å, which is slightly lower than the predicted value. However, the difference of 10 – 30 meV is considered a normal experimental error corresponding to relatively high accuracy. In 2011, Debbichi et al. investigated the electronic and optical properties of the short-period InAs/GaSb/InSb SLs grown on GaSb substrates by the modified 8×8 EFA model that took into account the effects of anisotropy and interfacial interactions [50]. The obtained results revealed that the bandgaps calculated at different temperatures were in good agreement with experimental data, which confirmed the high accuracy of the utilized model. In 2013, Yi Zhou et al. used the InAs/InAsSb/GaSb/InAsSb four-layer SL (including the interface) to replace the standard InAs/GaSb double-layer structure in the EFA model and optimize the n ML InAs/12 ML GaSb SL band structure [51]. The obtained fitting data presented in Fig. 13 indicate that the cut-off wavelength error of the four-layered structure is less than 5%, which is much lower than that of the standard SL model.

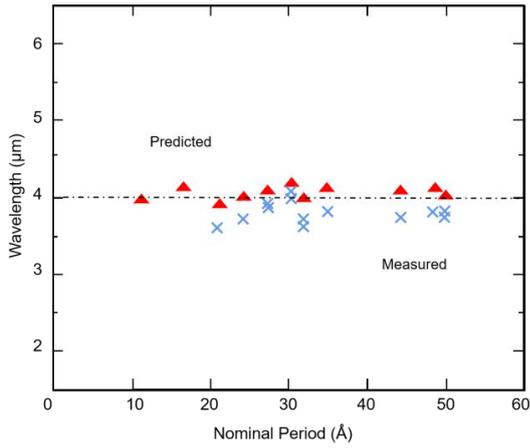

Fig. 12. Theoretical bandgaps of the modified EFA model (▲) and experimental PL peak energies (×) obtained at various nominal periods. The PL energy is close to the SL bandgap due to the small exciton binding energy [52].

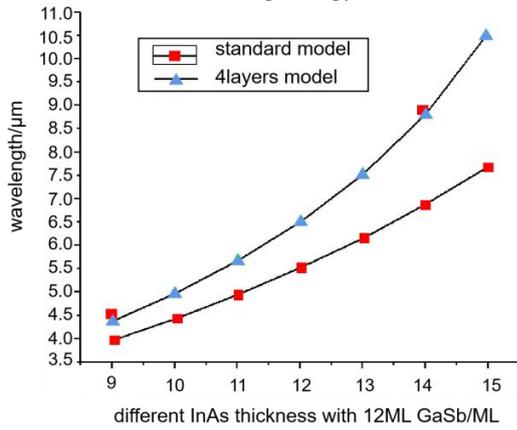

Fig. 13. Cut-off wavelengths of different SL structures calculated using the standard and four-layered EFA models, respectively. The circles corresponding to 9 and 14 ML of InAs represent the response wavelength positions obtained from the PL spectra of the laboratory-grown medium wave and longwave materials, respectively [53].

Boutramine et al. applied the EFA method to study the 25 Å InAs/25 Å GaSb SL bandgap as a function of the InAs layer thicknesses $d_1$ and temperature [54]. With increasing $d_1$, the electronic energy $E_1$ of the InAs layer decreased, and the heavy-hole energy $HH_1$ of the GaSb layer increased. This result was consistent with the values predicted using the k·p method [55]. As the temperature increased, the bandgap decreased from 288.7 meV at 4.2 K to 230 meV at 300 K. The corresponding cut-off wavelengths were equal to 4.3 and 5.4 µm, respectively, which belonged to the MWIR range. These parameters were in good agreement with experimental values [56]. In 2020, Benchtaber et al. calculated the band structure and bandgap of the 21 Å InAs/24 Å GaSb SL using the EFA model [57]. The bandgaps obtained at 5 and 300 K were equal to 316 and 247 meV, respectively, and the former value was consistent with the PL spectrum recorded at 300 meV [58]. The authors suggested that the small difference of 16 meV (0.05%) might be due to a valence band shift or the low thickness measurement accuracy. Boutramine, et al. also used EFA to do a lot of research [59; 60]on InAs/GaSb SL band structure, sub-bands and effective masses of carriers with different periods and the valence band offset Λ [61]. These findings are consistent with the experimental results reported by Cervera, et al. [56] . Hostut, et al. analyzed bandgap energy and hh – lh splitting energy of the InAs/AlSb/GaSb structure T2SL (N-type) using the EFA method [62]. As shown in Fig. 14, the bandgap of the structure was obtained as 144 meV, corresponding wavelength of 8.6 µm, which lay in the LWIR of the atmospheric window. And the hh–lh splitting energy was 166 meV, which was 22 meV higher than the bandgap. It is well known that the larger hh–lh splitting energy is very important for suppressing the Auger recombination and improving the minority carrier lifetime.

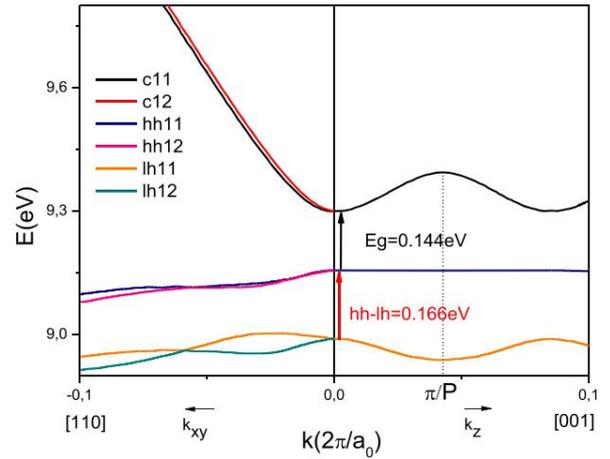

Fig. 14. Calculated band-structure of 13.5 ML/2 ML/8 ML InAs/AlSb/GaSb T2SL showing dispersion of energy with respect to electron wave vector in plane ($k_{xy}$) and in the growth direction ($k_z$) for LWIR T2SL. The in plane dispersion direction along [110] is shown on the left-hand, whereas the dispersion in the growth direction [001] is shown on the right hand. Conduction bands 1 and 2 zone edges at ( π /P) are presented as verticle dashed line (P is periodical length).

The EFA model requires a large number of numerical calculations and thus represents a complex theoretical approach. It is suitable not only for calculating the band structure of long-period SLs, especially the electronic states near the Γ point of the Brillouin zone (BZ), but also for heterostructural modeling. Some researchers have suggested that certain interface envelope function boundary conditions cause other uncertainties [20; 63]. For example, the EFA method produces similar results for different thin-layer SLs, which complicates the analysis and drawing conclusions from calculation data.

### 3.3 Empirical Pseudopotential Method

EPM is an atomic calculation method that is advantageous for studying short-period or thin-layer material structures such as semiconductors and metals. First, the model determines a SL form factor from a large number of band parameters. Alternatively, the bandgap obtained by fitting PL and light absorption spectra can be used to optimize the theoretically calculated SL bandgap [64]. EPM is considered a more accurate theoretical simulation method than the k·p and EFA models [25].

The famous scientist Fermi introduced a pseudopotential concept as early as in 1934 when he was studying the high-level electronic states. The pseudopotential approach was implemented for cases, in which the strong Coulomb potential in the vicinity of a nucleus led to the near-free electron approximation failure [22]. Among various pseudopotential techniques, the empirical pseudopotential method proposed by Phillips and Kleinman Phillips and Kleinman [65] in 1959 is the most representative one. Jianbai and Baldereschi developed an EPM method for simulating long-period SLs and successfully calculated the electronic structure of type I GaAs/AlGaAs SLs in 1987 [66]. They found that this method could also effectively model T2SLs. Miao et al. and Liu et al. used this technique to introduce an imaginary crystal Hamiltonian and accurately calculated the band edge structure of InAs/GaSb T2SLs in the (001) direction with different thicknesses of the InAs and GaSb layers in the 1990s [67]. They concluded that the SL conduction band edge mainly depended on the InAs layer thickness and that the valence band edge was mainly affected by the thickness of the GaSb layer, which provided a basis for the band structural design of T2SLs.

The atomic empirical pseudopotential method (AEPM) was originally proposed by Dente and Tilton in 1999 [68]. It uses the exact superposition of atomic pseudopotentials to calculate the energy and wavefunctions by solving a block equation for each atom and thus can simulate the interfacial structure very accurately. The authors successfully applied this technique to match the bandgap of the InAs/GaSb SLs reported in the study of Ram-Mohan[69] with an experimental PL spectrum. Finally, the AEPM was applied to the LWIR W-type laser AlSb/AlAsSb/InAs/GaSb/InAs SL structures with different InAs layer thicknesses. The cut-off wavelengths obtained from the corresponding PL spectra were 3.40, 3.85, and 4.40 μm, while the calculated values were 3.53, 4.01, and 4.52 μm, respectively. Hence, both datasets were almost identical within an error bar.

Subsequently, Dente and Tilton replaced the SL block function with a SL pseudopotential using pseudopotential form factors [70]. This led to a simpler version of the AEPM named the superlattice empirical pseudopotential method (SEPM), which assumed the redistribution of charges at the heterogeneous interface, making the SL components as bulky as possible and the energy – as low as possible. The authors found that the bandgaps of InAs and GaSb bulk materials were equal to 0.368 and 0.8 eV at 77 K, which slightly differed from the actual values of 0.371 and 0.8 eV, respectively. This confirmed the SEPM ability to accurately model SL electronic structures. In 2013, Masur et al. extended the standard two-component SEPM into a four-component model. That is, the structure uses a four-layer structure of InAs/InAsSb/GaAsSb/InAsSb including the interfaces as a full cycle [71]. As shown in Figs. 15(a) and (b), the two-component model is only suitable for SLs with layer thicknesses greater than 7 ML, whereas the four-component SEPM is significantly more accurate, and its calculated bandgaps closely match experimental values.

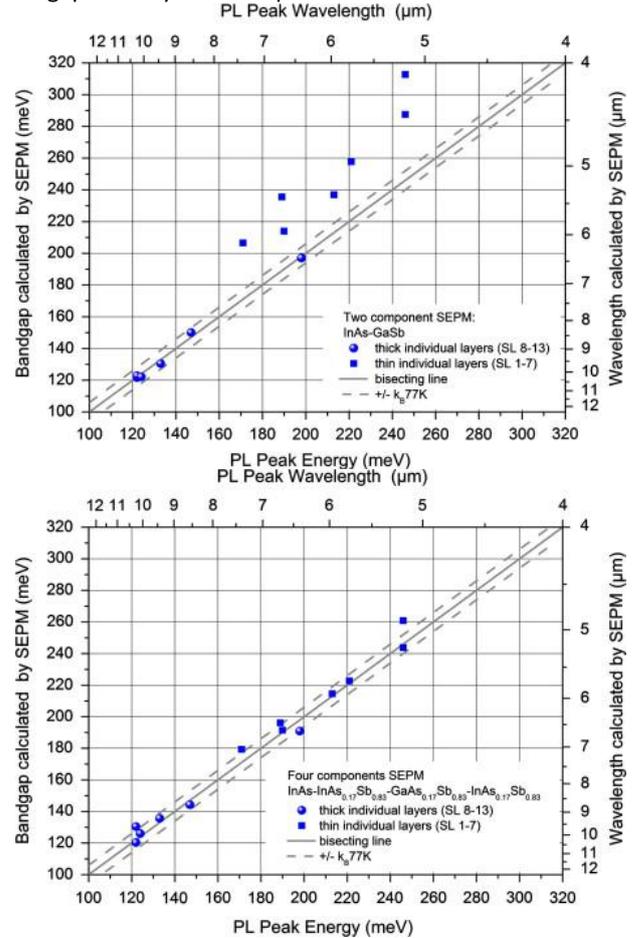

Fig. 15. Bandgaps calculated by the (a) two-component (InAs–GaSb) and (b) four-component (InAs – InAs$_{0.17}$Sb$_{0.83}$ – GaAs$_{0.17}$Sb$_{0.83}$ – InAs$_{0.17}$Sb$_{0.83}$) SEPMs and their experimental values obtained from the results of PL measurements conducted at 10 K for T2SLs with variable layer thicknesses. The circles and square represent thick individual layers (SL samples ≧ 7 MLs) and thin individual layers (SL samples ≦ 7 MLs), respectively.

In 2001, Ongstad et al. calculated the band structure of InAs/GaSb SLs by EPM [41]. The obtained results showed that the peak wavelength of the obtained PL spectra varied from 4.2 μm for an 8 ML/8 ML sample to 3.35 μm for an 8 ML/40 ML sample, leading to a blue shift. The EPM data were in good agreement with experimental absorption and PL spectra, and the bandgap ultimately converged to a constant value. In 2003, Magri et al. adopted the EPM to change the composition of interfacial bonds by exchanging only one interfacial anion plane between Sb and As atoms and determine the relationship between the bandgap and interfacial composition [72]. They theoretically predicted the separation interface blueshift of 64 meV and mutation interface blueshift of 95 meV (the corresponding experimental value was 70 meV). This result indicates that EPM is suitable only for the SL separation interface. The same research team published another study that compared several EPM methods (Fig. 16). Here, the bandgap calculated by AEPM was much smaller than that determined by the Dente and Tilton EPM method [71], which was close to the experimental value [73]. Piquini et al. used EPM to simulate the band edges and bandgaps of n ML InAs/m ML GaSb SLs for GaSb and InAs substrates, $C_{2v}$ and $D_{2d}$ dot groups, and (001) and (110) growth directions, respectively, and compared them with experimental data [74]. The obtained results revealed that EPM could accurately predict the energy band structures of T2SLs containing thin layers.

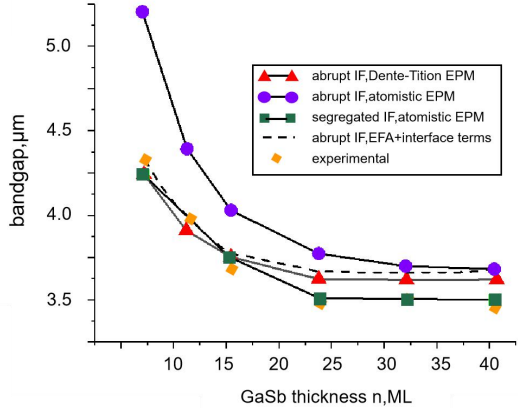

Fig. 16. Bandgaps of the $(InAs)_8/(GaSb)_n$ SLs predicted by the EPM developed by Dente et al., EFA model containing interfacial terms, and proposed method for abrupt and segregated interfaces.

In 2015, Çakan et al. used EPM to calculate the band structures of InAs, GaSb, and GaAs interfaces as well as InSb interface under strain [75]. To verify the obtained EPM data, the latter were compared with the results obtained by a hybrid density functional theory (DFT) HSE method and experimental bandgaps. The EPM and measured bandgaps were exactly the same and equal to 0.41, 0.81, 1.51, and 0.23 eV [76]. The bandgap values calculated by the HSE method were 0.34, 0.81, 1.36, and 0.27 eV, respectively, which significantly differed from the experimental ones. In 2018, Akel et al. simulated the energy band structure of InAs/GaSb and InAs/AlSb/GaSb N-type T2SLs by EPM to determine the dependences of the SL bandgap and hh–lh splitting energy on the AlSb/GaSb and InAs layer thicknesses [64]. Note that the hh–lh splitting energy, which is larger than the bandgap, can effectively suppress the Auger recombination process. The best results with a minimum bandgap of 128 meV and hh–lh splitting energy of 194 meV were achievrf for the 17 ML InAs/3 ML AlSb/6 ML GaSb SLs. Fig. 17 shows the bandgaps of two T2SL structures, x ML InAs/3 ML AlSb/6 ML GaSb and x ML InAs/9 ML GaSb, which decrease with an increase in the InAs layer thickness. At x = 6, the bandgaps and the corresponding wavelengths were equal to 354 meV and 3.5 μm, and 248 meV and 5 μm, respectively, which were within the MWIR range. Hence, this study accurately calculated the bandgaps and hh–lh splitting energies of the MWIR and LWIR bands, which could be potentially used for designing photodetectors operating in these ranges. 2021 Akel, et al. calculated the interband optical absorption of the InAs/GaSb T2SL structures [77]. The EPM method prediction showed that the SLs bandgap was underestimated about 0.4 μm, which corresponded to an uncertainty of less than 0.3 ML in the layer width, and it was in full agreement with the findings of Livneh et al.[23], Hostut and Ergun applied the method to calculate the band structure of the InAs/GaSb based T2SL [78]. The energy gap was measured as 246 meV with only ±10 meV variation compared to the experimental result at $\Gamma$ point (k = 0).

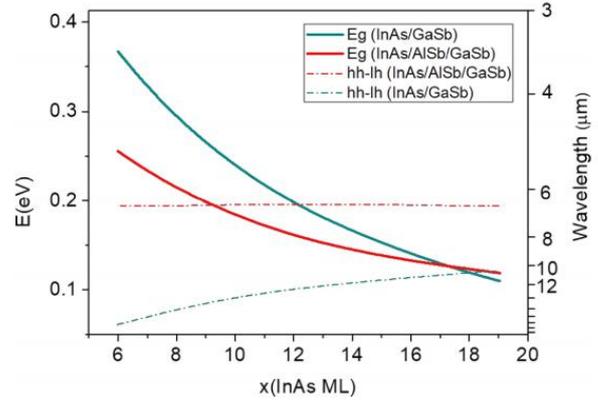

Fig. 17. $E_g$ and hh–lh splitting energies plotted as functions of x (InAs ML) for the $(InAs)_x/(GaSb)_9$ and $(InAs)_x/(AlSb)_3/(GaSb)_6$ T2SL structures with x varying from 6 to 19 ML. when InAs thickness is 17 ML, $E_g$ values intersect at 124 meV with corresponding wavelength of 10 μm within the long wavelength range. Bandgap energy decreases below the hh-lh splitting energy as InAs ML increases.

It is noteworthy that EPM is a non-self-consistent atomic method that takes into account interfacial effects. The greatest advantage of this technique is its simple form, which is easy to implement and allows calculating the energy bands of any crystal with minimum manpower and material resources. SEPM exhibits high calculation efficiency, and the obtained wavelength tuning data for type II antimonide lasers are in good agreement with experimental values [79]. However, EPM method cannot solve application problems in different chemical environments, especially when dealing with extremely delicate situations such as the charge transfer near the interface of a thin layer.

### 3.4 Empirical Tight-bound Method

When studying SL properties, ETBM not only considers the effects of strain, interface, and antimony segregation but also uses first principles. This model can estimate specific bandgaps according to the requirements of heterostructural calculations. Thus, it may reproduce important band structural features better than the standard model and is suitable for modeling quantum well heterostructures [80].

ETBM was initially proposed by Bloch in 1929 and subsequently used to determine the periodic potentials of solids by Slater and Koster in 1954 [81; 82]. It was originally called a linear combination of atomic orbitals. In 1983, P. et al. proposed a nearest-neighbor semi-empirical tight-binding theory of sphalerite materials [83]. They assumed that the ETBM model could solve the problem of material variation at the atomic scale and retain the complete crystal and electronic symmetries of semiconductor materials. This theory was eventually applied to studying bulk materials such as GaSb and InAs.

In 2000, Klimeck et al. fitted the orbital interaction energies of nine binary compounds, including InAs and GaSb, with the sp3s* ETBM model at room temperature using literature data [83; 84] as target values [85]. The calculated energies of the lowest conduction bands of InAs and GaSb were 0.368 and 0.751 eV, while the corresponding target values were 0.370 and 0.750 eV, respectively. The calculated energies of the three highest valence bands of InAs and GaSb were −12.159, 4.126, and 4.543 eV and −12.683, 3.123, and 4.033 eV with the corresponding target values of −12.300, 4.390, and 4.630 eV and −12.000, 3.400, and 4.700 eV, respectively. There results indicate that the utilized method is very effective in predicting bandgaps. Subsequently, Wei and Razeghi modeled InAs/GaSb SLs with InSb interfaces using the more accurate sp3s* ETBM by considering the antimony bias in the InAs layer [86]. The authors computationally fixed the thickness of the GaSb layer at 40 Å (13 ML) and varied the InAs layer thickness from 40 Å (13 ML) to 66 Å (22 ML). The calculated bandgaps were compared with experimental values. As shown in Fig. 18, the experimental datapoints are closely scattered around the calculated curve withing an uncertainty range. This confirms that ETBM, which considers the interfacial and antimony segregation effects, is a reliable method for the SL design process.

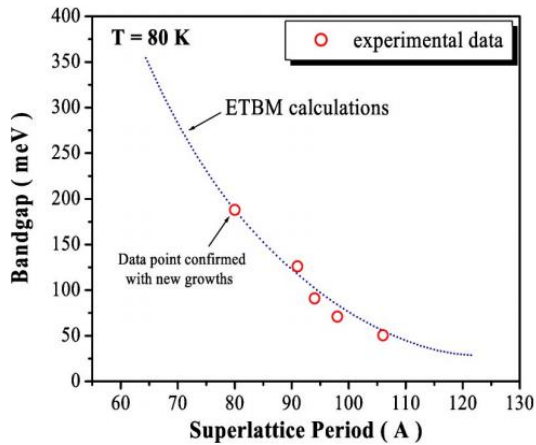

Fig. 18. Bandgap values calculated by the modified ETBM and the corresponding experimental data by fixing the GaSb layers at 40 Å (13 ML), and varying the thickness of the InAs layers from 40 Å (13 ML) up to 66 Å (22 ML). The experimental data points scatter closely round the calculated curve. The cutoff wavelength for the superlattice with an InAs layer thickness of 40 Å has been confirmed with newer growths using the same growth conditions.

The concept of M-type structure was initially proposed by Nguyen et al. for the InAs/GaSb and InAs/GaSb/AlSb/GaSb SLs in 2007 [87]. It maintains the type II band arrangement, which can significantly reduce the Auger transition of the electron–hole splitting band, while the AlSb barrier decreases the dark current serving as a barrier. Using this approach, the energies of the valence and conduction bands may be significantly adjusted. The authors calculated the electronic energy bands of such structures by ETBM and grew six M-type SL structures with different layer thicknesses by molecular beam epitaxy under the same conditions. The obtained PL spectra were consistent with the theoretical energy gaps (Fig. 19), indicating that M-type structures could exhibit cut-off wavelengths greater than 11 μm. Nguyen et al. applied the above-mentioned sp3s* ETBM and determined the minimum and maximum values of the conduction and valence bands of the InAs/GaSb and M-type SLs by performing energy band structure calculations (Fig. 20), confirming the ability of this method to adjust the band edges of the M-type structure [88]. Razeghi and Nguyen also simulated the M-type structure of InAs/GaSb SLs using ETBM and concluded that the utilized method could effectively simulate the M-type structure of type II SLs [89]. Czuba, et al. also used the sp3s* tight-binding model to calculate the electronic structure of the inter-band cascade infrared detector with a cut-off wavelength of 10.7 μm, and obtained its effective bandgap [90]. In 2021, Zhu, et al. considered the thermal strain in the sp3s* ETBM method and calculated the energy band structure of InAs/GaSb T2SLs devices. The measured PL spectrum and spectral responsivity verified the simulation results [91]. In the device design stage, the performance of the device can be simulated to guide the design and achieve the required performance.

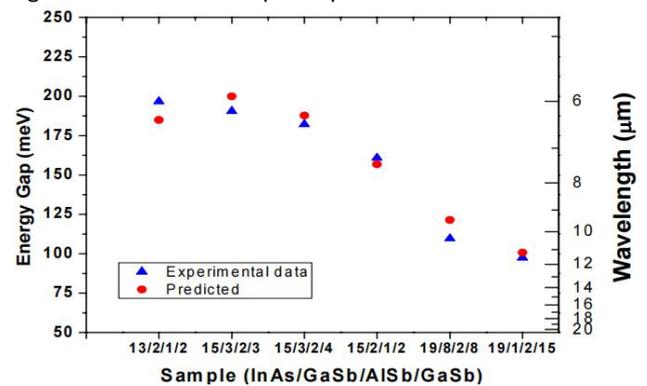

Fig. 19. Theoretical energy gaps and experimental photoluminescence peaks obtained for some M-type structures. The theoretical model and experiment are a good agreement. And it realizes the M-structure with cut-off wavelength beyond 11 μm.

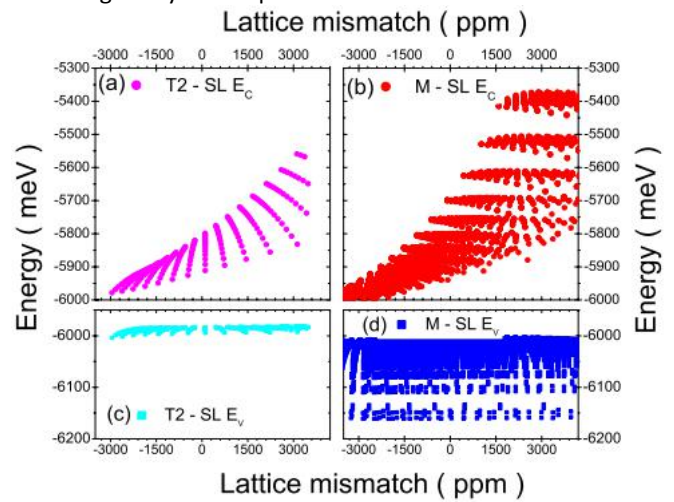

Fig. 20. Energy distributions of various SL designs: (a) type II InAs/GaSb conduction band, (b) M-type structure conduction band, (c) type II InAs/GaSb valence band, and (d) M-type structure valence band. The M-type structure exhibits a larger bandwidth.

In 2018, Akitaka et al. determined the parameters of the sp3d5s* ETBM model established for InAs and GaSb semiconductors and used the sp3s* and sp3d5s* models to fit their bandgaps calculated by the hybrid QSGW and ETBM approaches [92]. As seen in Fig. 21, the lowest valence band energy determined the sp3s* ETBM is almost flat between points X, W, U, and K, while the sp3d5s* method successfully solves this problem and matches the results of hybrid QSGW calculations. This shows that the sp3d5s* EBTM model significantly enhances the sp3s* model and is suitable for guiding the SL design process.

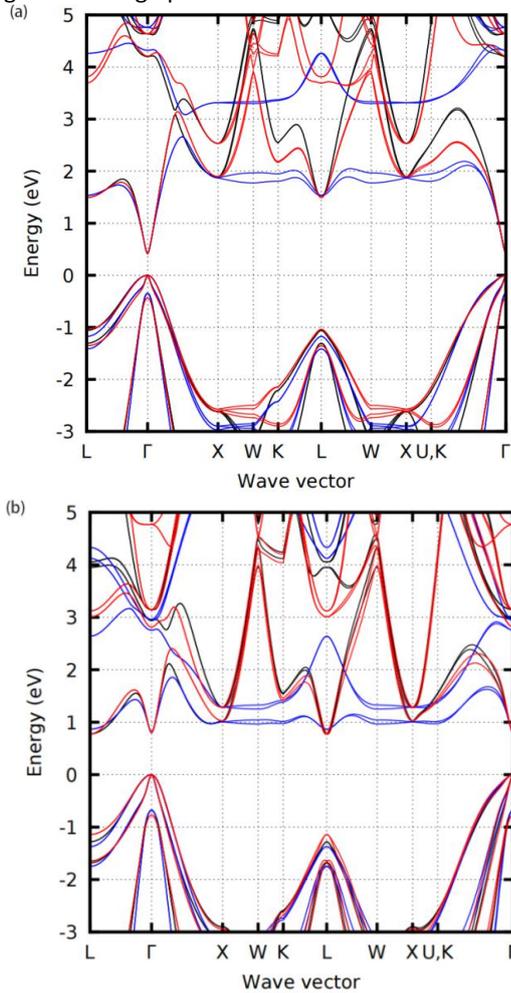

Fig. 21. Band structures of (a) InAs and (b) GaSb obtained by the hybrid QSGW (black) and ETBM calculations (the blue color denotes the sp3s* model, and the red color represents the sp3d5s* model). Whereas the lowest conduction band is fitted moderately well by the sp3s* model along the left two segments (L→ Γ →X), but it is almost flat between the X, W, and U, K points, reflecting the poorly described transverse mass. The sp3d5s* method resolves this problem.

In 2017, Jiang, et al. reported the flexibility of adjusting the valence band level by inserting a thin (0.6 ML) InSb layer in the middle of the GaSb layer of 15 ML InAs/7ML GaSb T2SL [93]. They used the ETBM method to calculate the tunability of the bandgap in the VLWIR SLs, and the results showed that this method extended the cut-off wavelength from 14.5μm to 18.2 μm. The consistency of theoretical prediction and experimental measurement shows that this advantage can be used to achieve very long-wave infrared detection without increasing the thickness of the InAs layer. Nejad and Sheshkelani modeled the 15 ML InAs/4 ML AlSb T2SL short-wave infrared detector using the ETBM model and proposed the corresponding band structure extraction algorithm in 2020 [94]. The result showed that the ETBM simulated cut-off wavelength differs from the experiment about 30 nm, which was a good match. It is an accurate method to model the SWIR T2SL bandgap.

Compared with other methods, ETBM is advantageous for calculating the structure of the entire BZ and accurately describes the parameters of various structures from atoms to SLs. It also considers material imperfections during growth, does not require a large number of complex numerical calculations, and is relatively fast. However, proper parameter selection is very important for its practical implementation. The overlapping parameters have a clear and simple physical meaning, and their number increases rapidly with an increase in the number of neighbors, which negatively affects computational accuracy.

### 3.5 First-principles calculations
#### 3.5.1 Density Functional Theory

The DFT was developed by Hohenberg, Kohn, and Sham in the early 1960s [95; 96]. It has been used as a standard approach for calculating the electronic structures of solids. Typically, the bandgap of a SL has a small positive value. However, DFT and other DFT-based first-principles calculation methods such as local density approximation (LDA) and generalized gradient approximation (GGA) often produce zero or even negative bandgap values due to the presence of a non-physical Coulombic self-repulsion term, which leads to a systematic bandgap underestimation. Because DFT methods are fast and widely available, many researchers have attempted to improve them to solve this problem.

In 2012, Sun and Zheng used the GGA method involving all-electronic relativity to calculate the Γ-point bandgaps of InAs, GaSb, GaAs, and InSb semiconductors related to InAs/GaSb SLs [97]. The obtained results showed that for InAs and InSb with very small bandgaps, the computed values amounted to 0 eV, while the calculated bandgaps of GaAs and GaSb equal to 1.258 and 0.603 eV, respectively, were much smaller than experimental values. Thus, the electronic bandgaps calculated by the GGA method are generally lower than experimental values, and the calculation accuracy depends on a particular compound.

Caid et al. studied InAs and GaSb compounds using a full-potential linear muffin-tin orbital approach based on LDA in 2019 [98]. The obtained electronic band structure was in good agreement with the results of previous calculations, and the corresponding bandgap was almost 0 eV. Because the corresponding experimental values were 0.42 and 0.72 eV, respectively, the LDA method significantly underestimated the bandgaps of these compounds.

In 2006, Becke and Johnson proposed a very simple and effective local potential that considerably increased the bandgap computational accuracy [99]. Fabien and Peter developed a modified Becke – Johnson potential combined with a local density approximation (MBJLDA) method for bandgap calculations in 2009 [100]. As shown in Fig. 22, the bandgaps calculated by the MBJLDA technique much better matched experimental values than the magnitudes computed by the LDA, hybrid functional, and GW methods [101; 102]. Subsequently, Kim et al. obtained the conduction and valence bands of five semiconductor bulk materials, including InAs and GaSb, at their Γ, X, and L points by the MBJLDA, GW, and hybrid generalized methods [103]. They found that this MBJLDA method with a modified potential produced more accurate band topologies and bandgap values than the other methods. In 2016, Gmitra and Fabian used the first-principles full-potential linear augmented plane wave method to reproduce the experimental bandgaps of GaSb and InAs [104]. In that study, the TB–MBJ exchange potential was used as the exchange energy term. The calculated bandgap was 0.822 and 0.417 eV, and the corresponding experimental value was equal to 0.812 and 0.417 eV, respectively [76]. Hence, the calculated values were consistent with the experimental ones within a certain error. The authors also compared their data with the bandgap computed by Chantis et al. [105] using the GW method, which were equal to 1.16 and 0.68 eV, respectively. Thus, it can be concluded that the TB – MBJ exchange potential produces more accurate bandgap values than other DFT techniques. In recent years, Patra, et al. used LDA and MBJLDA to perform theoretical analysis and prediction of superlattice heterostructures, verifying that change in the thickness of the InSb layer could lead to the change between the direct and indirect bandgap[106]. Furthermore, the MBJLDA method is reliable and may provide a benchmark for the empirical fitting of energy band structures.

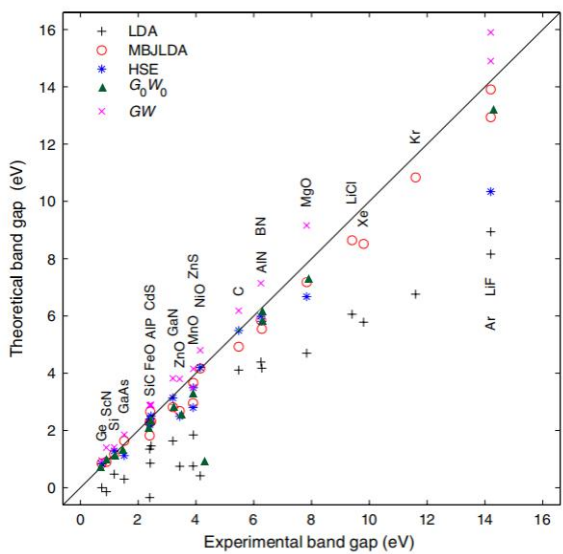

Fig. 22. Theoretical and experimental bandgaps of 23 different solid compounds calculated by various DFT methods, include the LDA, MBJLDA potentials and other methods (HSE, all-electron non-self-consistent $G_0W_0$, and self-consistent GW). For most cases, the MBJLDA potential yields band gaps which are in good agreement with experiment leading to typical errors of less than 10%.

In 2014, Wang and Zhang attempted to predict the electronic structure of GaSb/InAs SLs by first-principles methods [39]. They empirically corrected the s, p, and d components of the atomic pseudopotential constants by performing self-consistent calculations and adjusted the DFT electronic structures of GaSb and InAs to match experimental values or quasi-particle calculation data and thus correct the bandgap error caused by the LDA method. Subsequently, the authors studied n/8 ML SLs and compared the calculated values with experimental and EPM data to verify their accuracy. As shown in Fig. 23, although the calculated values are slightly higher than the experimental ones, they are still closer to the measured values than the magnitudes obtained by the EPM model.

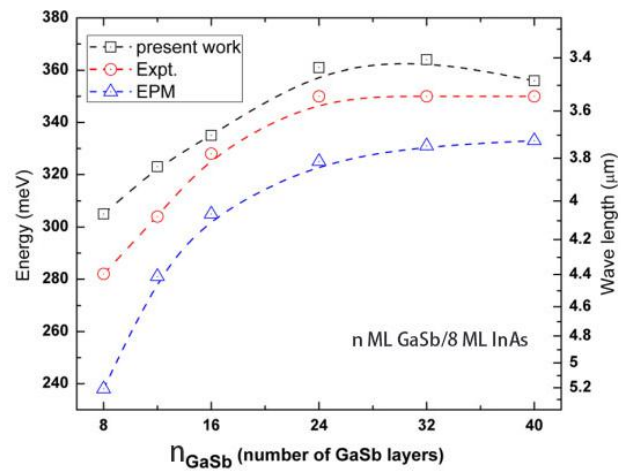

Fig. 23. Transition energies from the first valence band to the first conduction band determined by various methods. The black squares indicate calculated results by empirically corrected LDA method, the red circle the experimental values, and the blue triangle the results calculated by EPM. LDA values are closer to the experimental data than the EPM results. The inter-layer atomic diffusion tends to lower the bandgap from that predicted assuming abrupt interfaces, so the calculated values are slightly higher.

Castaño-González et al. studied the electronic structure properties of GaSb using traditional DFT methods, including LDA, GGA (PBE, PBEsol, PW91, rPBE, AM05), and meta-GGA (TPSS, RTPSS, MBJ) techniques [107]. Their work provided a reference for selecting the most suitable DFT method. It was found that the bandgaps obtained by the LDA and GGA approaches were close to zero. Meanwhile, the values calculated by the MBJ, and RTPSS methods based on the meta-GGA functional and AM05 method using the GGA functional were equal to 1.015, 0.513, and 0.482 eV, respectively. These magnitudes were very close to the experimental values reported in the literature, and the MBJ and RTPSS techniques produced the lowest relative errors.

The hybrid functional incorporates a part of the Hartree–Fock exchange energy term to solve the bandgap problem. In 2010, Tomić et al. studied the hybrid exchange generalized functions B3LYP and PBE0, which contained three widely used generalized functions (LDA, GGA, and PBE) [108]. As shown in Fig. 24, the PBE0 method overestimated the bandgaps, while the B3LYP approach produced slightly more accurate data. In 2016, Garza and Scuseria predicted the bandgaps of InAs and GaSb bulk materials and compared the performances of four commonly used hybrid density functionals (HSE, B3PW91, B3LYP, and PBE0) [109]. As shown in Fig. 25, these four methods somewhat overestimate the bandgap values; however, the obtained results are relatively concentrated, and the HSE, B3PW91, and B3LYP data are closer to the experimental values within an error. Moreover, although PBE0 overestimates the bandgap, the error is systematic and can be corrected by linear fitting. Considering the cost and accuracy of this method, it may potentially have a wide application range. Yao et al. calculated the electron band structures of InAs/GaSb SLs in the (111) direction and compared the results of hybrid functional calculations with those obtained by the ordinary DFT methods [110]. The bandgaps of InAs and GaSb were also determined by the traditional PBE method and hybrid functional HSE–PBE, HSE–PBEsol, B3LYP as well as experimentally [111]. The obtained magnitudes were 0.00, 0.36, 0.45, 0.43, and 0.42 eV (for InAs) and 0.00, 1.02, 0.99, 0.34, and 0.81 eV (for GaSb). Hence, the traditional PBE method cannot accurately predict the bandgaps of InAs and GaSb, while the three hybrid functional methods produced relatively narrow bandgaps of these two compounds (the best results were obtained by the HS–PBEsol approach). Garwood, et al. calculated the bandgaps of InAs/GaSb T2SLs using the PBE0 hybrid functional, which was consistent with the calculated value using the 18% exact exchange function, and the deviation range from the experimental bandgap was 3%-11%[112]. In 2021, Yang, et al. applied Bayesian optimization (BO) machine learning to introduce a new method of DFT with Hubbard U correctionthe calculational methods [113]. According to report [114], InAs and GaSb the optimal values were showed as follow: $U_{eff}^{In,p} = -0.5 eV$ , $U_{eff}^{As,p} = -7.5 eV$ , $U_{eff}^{Ga,p} = 0.8 eV$ , $U_{eff}^{Sb,p} = -6.9 eV$ .They compared the band structures calculated by PBE+(BO) and HSE. It could be seen from Fig. 26 that PBE+U(BO) and HSE are generally in good agreement, but PBE+U(BO) underestimated the bandgap.

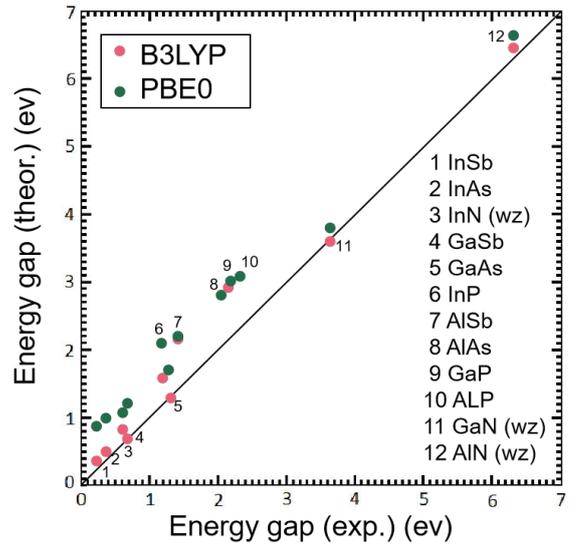

Fig. 24. A comparison of the experimental energy gaps with those predicted using the DFT with B3LYP, PBE0, and DFT functionals and experimental lattice constants. The B3LYP functional provides slightly better agreement with the experiment then PBE0, And the PBE0 functional results in an overestimate of the band gap.

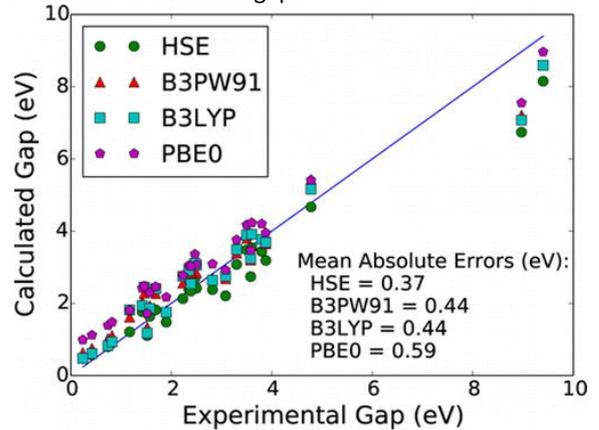

Fig. 25. Calculated and experimental bandgaps determined for various compounds using DFT functionals. B3LYP and B3PW91 give extremely similar band gaps. This is most likely due to the incorporation of similar amounts of HF exchange, B3LYP and B3PW91 both have 20% full-range nonlocal exchange, while HSE includes 25% but only in the short-range, this reason also increases the calculated band gaps.

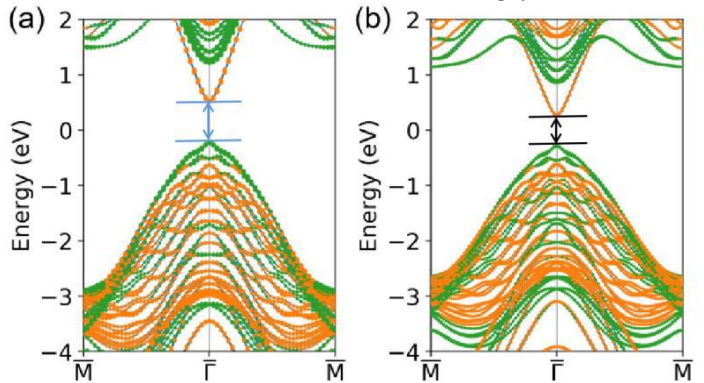

Fig. 26. The band structure of an InAs/GaSb interface with 5 layers of InAs and 5 layers of GaSb obtained with (a) HSE and (b) PBE+U(BO). Orange and green dots indicate the contributions of InAs and GaSb, respectively. PBE+U(BO) somewhat underestimates the band gap.

Asadi and Nourbakhsh calculated the bandgap of InAs using the Trouiller–Martins pseudopotential combined with LDA (FHI.LDA), Hartwigsen – Goedecker – Hutter pseudopotential with parametric conservation (HGH.G0W0), and the FHI.G0W0 method including the In 4d state [115]. The calculated and experimental bandgap values were 0.181, 0.845, 0.483, and 0.417 eV, respectively, indicating that the FHI.G0W0 data obtained using the multi-body perturbation theory were in good agreement with the experimental results. In 2020, the same studies the electronic structure of InAs in the DFT framework for the first time, using LDA and the norm-conserving pseudopotential that treated the In 4d electrons as valence electrons [116]. The computed bandgap without spin coupling was 0.368 eV, which underestimated the experimentally determined value of 0.420 eV by only 12.38%.

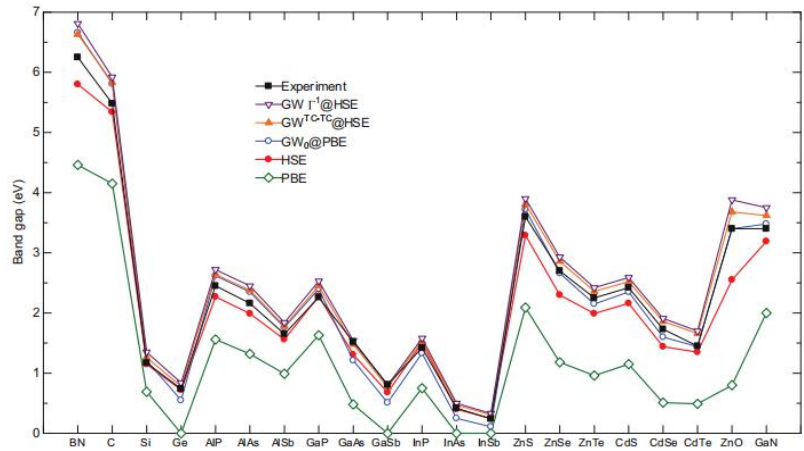

Fig. 27. Bandgaps obtained by various DFT approximations (PBE, HSE, $GW_0$@PBE, $GW^{TC-TC}$@HSE, and $GW\Gamma^{-1}$@HSE) and the corresponding experimental values. PBE semilocal functional has the largest deviation. The three fitting results related to GW method are consistent with experimental band gap values.

In general, the DFT approach does not rely on empirical or experimental parameters. However, DFT methods typically require performing a large number of calculations when simulating nanostructures composed of thousands atoms, leading to very small bandgap values. Several methods mentioned above were able to significantly improve the prediction accuracy of the electronic structures and bandgaps of various semiconductors. They included the TB – MBJ exchange potential with a localized potential, pseudopotential correction, treatment of the In 4d electrons in pseudopotentials as valence electrons, and hybridized generalization techniques. However, the problem of bandgap underestimation by DFT methods has not been completely resolved yet.

**3.5.2 Many-body Perturbation Theory (GW Method)**

The GW method based on the many-body perturbation theory can be considered a DFT method with some modifications and perturbations, which make it suitable for calculating the excited states of a multi-body system. The widely used GW techniques include hybrid quasi-particle self-consistent GW (QSGW) and fast non-self-consistent GW (G0W0) methods. They are able to reduce the self-interaction error and solve the problem of very low theoretical bandgaps obtained by DFT methods due to the utilization of the Green function that accurately calculates the excitation spectra of quasiparticles, while the Kohn–Sham (KS) equation employed by DFT is not strictly physical.

The GW approximation was proposed in 1965 by Hedin to describe the kinetic response of a system to external perturbations [117]. The authors also derived a set of more accurate self-consistent equations containing the Green function. This step was critical for describing the quasiparticles of many-body systems and allowed more accurate prediction of the energy band structures of solids. In 2014, Hinuma et al. performed first-principles calculations using the PBE semi-local approximation, HSE hybrid approximation, and GW approximation to study the band arrangements of various semiconductors in the sphalerite structure [111]. They compared the bandgaps computed by the GW0@PBE, GWTC–TC, and GWΓ1 methods with the PBE, HSE, and experimental values. Fig. 27 shows that 1) GWΓ1 most accurately reproduces the experimental data; 2) the HSE approximation can reasonably predict bandgap energies; and 3) all three GW approximations produce small deviations from experimental values. In 2014, Kotani developed the PMT – QSGW method based on the hybrid all-electron total potential method (PMT), which used both augmented plane waves and muffin-tin orbitals [118; 119]. They studied the dependence of the GaAs bandgap on the number of k points in the first BZ for self-consistent energy calculations. As shown in Fig. 28, the GaAs bandgap at the Γ point smoothly converges with increasing number of k points, and the value obtained for the 4 × 4 × 4 k-point mesh is only 0.1 eV higher than that determined using the 10 × 10 × 10 mesh. These results can help select an optimal number of k points to maximize the computational accuracy at limited resources.

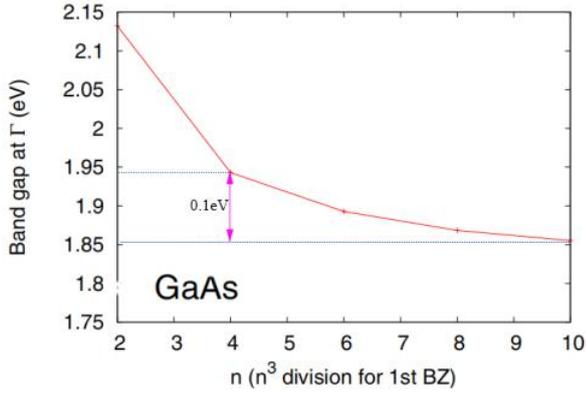

Fig. 28. GaAs bandgap dependence on the number of k points in the first Brillouin zone for self-consistent energy calculations. The integer n of the x-axis means that the number of divisions of BZ is n×n×n. The y-axis is for the band gap.

The computational cost of the complete GW calculation is too high, which led to the development of the fast non-self-consistent GW approximation (G0W0). In 2013, Malone and Cohen used the first-principles plane-wave pseudopotential and G0W0 methods to calculate the quasiparticle band structures of InAs and GaSb semiconductors [120]. The bandgap values Γ 15v and Γ 1c of the InAs bulk material computed at the Γ point by the LDA plane-wave pseudopotential, G0W0, and G0W0+SO methods, and their corresponding experimental magnitudes were equal to 0 and −0.46 eV, 0 and 0.55 eV, −0.38 and 0.42 eV, and −0.37 and 0.36 eV, respectively. The bandgap values of the GaSb bulk material were 0 and −0.13 eV, 0 and 0.94 eV, −0.73 and 0.70 eV, and −0.76 and 0.81 eV, respectively. Hence, the G0W0+SO data most closely matched the experimental values.

In 2016, Deguchi et al. proposed a hybrid QSGW method that consisted of 80% QSGW and 20% LDA and used it to calculate the band structures of various compounds, including InAs and GaSb [121]. The authors also compared the bandgaps estimated by the LDA and LDA+SO (0.00 eV), QSGW (0.8 and 1.2 eV), QSGW+SO (0.68 and 0.99 eV), QSGW80 (0.48 and 0.99 eV), and QSGW80+SO (0.36 and 0.77 eV) methods with the experimental values of 0.42 and 0.82 eV, respectively. The hybrid QSGW80+SO technique produced the closest match to the experimental data, indicating that it was one of the most accurate first-principles methods. Subsequently, Otsuka et al. applied this method for simulating a short-period InAs/GaSb infrared sensor [122]. They performed self-consistent calculations using the 4 × 4 × 3, 4 × 4 × 2, and 4 × 4 × 1 k-point meshes and calibrated the obtained bandgaps. Fig. 29 plots the bandgaps calculated by the QSGW, EPP [74] and ETB [86] methods at the Γ points and the corresponding experimental values derived from PL spectra [41] as functions of n. The bandgap energiess calculated by the QSGW method are in good agreement with the ETB results. This result was in full agreement with the comparison result of the empirical sp3s* tight-binding method (TB) and eight-band k · p method, the empirical pseudopotential method (EP), the QSGW method by Kato and Souma in 2018 [123]. Although the cited research study focused on small SLs, its findings can be used as a basis for applying ETB methods to larger SLs.

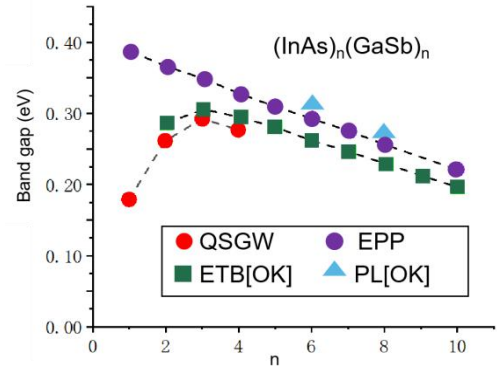

Fig. 29. Bandgaps of the n ML InAs/n MLGaSb SLs calculated by the QSGW80+SO (closed circles), EPP (open circles), and ETB (open squares) methods with the corresponding experimental values (closed triangles). The values obtained by the ETB and PL methods are extrapolated to those at zero temperature.

Considering the importance of the InSb interfacial layer, Taghipour et al. first applied the GW approximation to study the electronic structure of long-period InAs/GaSb T2SLs in 2018 [124]. Fig. 30 shows that DFT−LDA underestimates the bandgaps of all structures, while the bandgaps determined by the G0W0 method are close to the experimental values [125] and slightly differ from the GW0 data. Although the bandgap decreases with increasing thickness of the InAs layer, the simulation results do not follow this trend due to calculation errors. The electronic bandgaps of T2SLs predicted by the GW methods are relatively consistent with the experimental data.

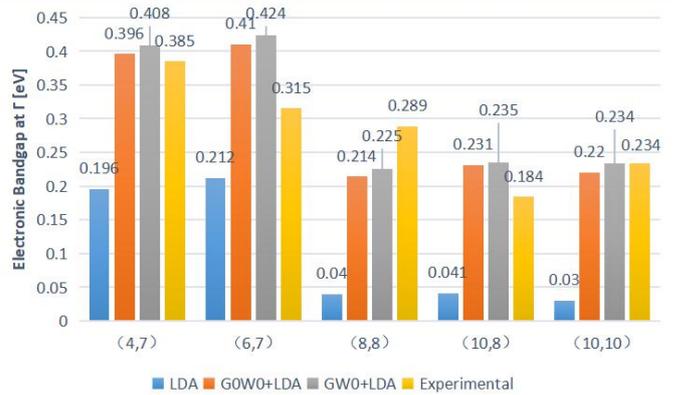

Fig. 30. Simulated bandgaps of (4,7), (6,7), (8,8), (10,8), and (10,10) InAs/GaSb SLs in the first Brillouin zone using the DFT − LDA, G0W0+LDA, and GW0+LDA approximations with the corresponding experimental values. GW+LDA lead to results in good agreement with the experiment.

Multi-body interactions are included into the GW methods to modify the calculated LDA energy band by considering the dynamic shielding exchange, self-energy caused by Coulombic holes, local field, and dynamic shielding effect. As a result, these techniques produce more accurate bandgaps in the calculations of semiconductor electronic band structures. The number of computational steps performed during these calculations is very large; therefore, all GW methods are relatively expensive and even more expensive than the hybrid generalized function approximation. Furthermore, their SL periods are limited by a value much lower than that of the real sensor, which is a major disadvantage of these techniques.

## 4 Method Limitations and General Outlook

As the most promising materials for infrared lasers and detectors, antimonide type II SLs have very impressive application prospects. The key to the rapid advancement of type II SL technology is the realization of the importance of the design and prediction steps before material growth. In particular, designing and predicting the electronic structures and bandgap of a material quickly and accurately is a time-saving and economical approach for both theoretical researchers and experimentalists. In recent decades, significant progress has been made in the prediction of class II SL energy bands. However, type II SL technology has failed to reach a theoretical prediction level comparable to that of the mature HgCdTe technology, and the calculated bandgaps of InAs/GaSb SLs are not very accurate, which significantly limits their practical use.

From the data discussed above, the following conclusions can be drawn.
- First, the quantum confinement of electrons and holes in the InAs/GaSb system is only possible for thin semiconductor layers. Second, only thin InAs and GaSb layers may produce a sufficient wavefunction overlap; however, the k·p method cannot adequately describe the electronic structure of a thin SL. Third, the k·p method does not take into account the coupling of different sub-bands, and its application to InAs/GaSb T2SLs produces a significant error in a case of strong energy band coupling. In addition, the k·p method, EPM, and ETBM are still unable to completely the resolve the point defect and interfacial problems.
- EFA calculations are very complex. Although EFA considers interfacial effects, it ignores the difference between the block functions of two constituent materials in the center of the BZ at the interface, and the wavefunction is directly matched at the interface. In addition, the boundary conditions of the model may cause other uncertainties.
- The first principles use the basic laws of physics to solve Schrödinger's equation. However, there are strong interactions between electrons, and the exact solution of the Schrodinger equation cannot be obtained for complex InAs/GaSb T2SLs.    More advanced functionals, such as the hybrid density and meta-GGA ones, depend on KS orbitals. The excited states of these models are simply treated as the differences between the KS energy levels of quasi-particles, and the calculations of the excited states and optical properties produce large errors.

Interfacial effects that are generally ignored in the calculations of the InAs/GaSb SL energy bands will be the focus of future studies on bandgap prediction methods. The problems caused by electronic interactions and the first-principles underestimation of bandgap values remain important challenges in the field of energy band calculations. In the field of semiconductor materials science and engineering, it is a promising research direction that should adopt a suitable theoretical method for calculating the electronic structure bandgap and adjusting the InAs/GaSb SL components to avoid the formation of defective energy levels in the forbidden band and optimize the energy band structure. In addition, the use of experimental data for guiding the theoretical design process is also an important direction of T2SL energy band research studies. At present, experimental results are mostly used to verify the accuracy of theoretical calculations, and a large gap exists between the synergetic realization of these two aspects.

## 5 Conclusion

Type II superlattices has received a lot of attention due to special energy band structure and excellent device performance. InAs/GaSb type II superlattices is a broken-gap band structure with the ability to adjust the positions of the conduction and valence band edges independently, providing abundant operating space for antimonide superlattices to carry out band engineering design. The energy gap falls well within the infrared regime by simply changing the components of the material and the thickness of each thin layer to be tuned, which is a key in infrared detection application. Simulating the energy band structure of InAs/GaSb T2SLs by various calculation methods, changing the bandgap, can make the theoretical calculations and experimental results complement and confirm each other. As a result, the performance of antimonide infrared detectors has reached that of HgCdTe-based systems in a relatively short period and even surpassed it in some aspects. The review covers the energy band structure of InAs/GaSb T2SLs, several commonly simulation calculation methods. And on this basis, the development direction of more suitable calculation methods oriented to interface effects and electronic interactions is proposed, which provides a reliable reference for the simulations of different superlattices structures.

## Author Contributions

Shuiliu Fang: Conceptualization, Data curation, Writing - original draft, Writing - review & editing. Ruiting Hao: Funding acquisition, Investigation, Methodology, Supervision, Project administration. Longgang Zhang: Writing - review & editing. Jie Guo: Writing - review & editing. Wuming Liu: Funding acquisition, Investigation, Methodology, Supervision, Project administration.

## Conflicts of interest

There are no conflicts to declare.

## Funding

This work was supported by the National Key R&D Program of China under grants No. 2016YFA0301500, NSFC under grants Nos. 61835013, 61774130, Strategic Priority Research Program of the Chinese Academy of Sciences under grants Nos. XDB01020300, XDB21030300.

## Notes and references